\documentclass{eptcs}
\providecommand{\event}{TTC 2013}

\usepackage{oz2}
\usepackage{graphicx}
\usepackage{breakurl}

\title{Case study: Class diagram restructuring}
\author{K. Lano, S. Kolahdouz-Rahimi\\
Dept. of Informatics, King's College London, Strand, London, UK\thanks{Research supported by the HoRTMoDA EPSRC project}}
\def\titlerunning{Case study: Class diagram restructuring}
\def\authorrunning{K. Lano, S. Kolahdouz-Rahimi}

\begin{document}
\maketitle

\begin{abstract}
This case study is an update-in-place refactoring transformation on
UML class diagrams. Its aim is to remove
clones of attributes from
a class diagram, and to identify new 
classes which abstract groups of classes
that share common data features.

It is used as one of a general collection
of transformations (such as the removal of redundant inheritance, or multiple
inheritance) which aim to improve the
quality of a specification or design level
class diagram.

 The transformation is a typical example
of a model refactoring, and illustrates
the issues involved in such transformations. 
\end{abstract}

\section{Introduction}
\label{introsec}

Update-in-place transformations have
specific challenges which need to be
addressed by model transformation
tools:
\begin{itemize}
\item Establishing confluence
of the transformation: that it produces 
a unique (up to isomorphism) result from
a given source model.
\item Ensuring termination of
a fixed-point implementation strategy.
\item Ensuring control over the order of
rule applications, in order to optimise
some characteristics of the transformation.
\end{itemize}

We have previously used the core version
of this case study to compare model
transformation approaches, and it has
proved to be effective as a test of the
capabilities of approaches to specify and
implement update-in-place 
transformations requiring fixed-point
iteration of rules \cite{Lano12scp}. 
The transformation
also requires relative
prioritisation of rules, and fine-grained
control of the order of application of 
individual rules to elements of a model.

The core version of the case study 
concerns removal of attribute clones
assuming that
only single inheritance exists
in the model
(Section \ref{attclones}), this is then
extended in Section \ref{assocclones}
to deal with the case where multiple
inheritance is present.

 In comparison to the refactoring case
 study of \cite{Pere10}, the present case
 study is focussed upon the automated
 selection of refactoring steps, instead of
 interactive refactoring, and includes
 class diagrams with multiple inheritance.
 In addition, the evaluation criteria of
 \cite{Pere10} only cover a subset of the
 aspects which this case evaluates. 
 Since
 we are interested in automated
 refactoring, termination and confluence
 are specific evaluation criteria which we
 are interested in, and the quality of the
 refactoring (ie., minimising any size
 increase in the model) is also 
 emphasised.

\section{Core problem}
\label{attclones}

\begin{figure}[htbp]
\begin{center}\
\includegraphics[width=4.2in]{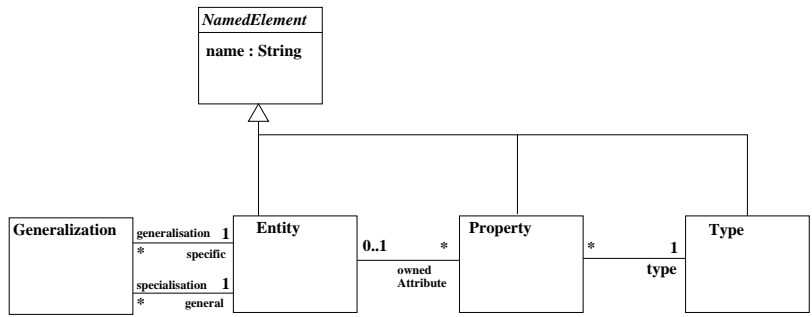}
\caption{Basic class diagram metamodel}
\label{cd0mm}
\end{center}
\end{figure}

Figure \ref{cd0mm} shows the metamodel for the source and target language of the
simple version (dealing with
single inheritance 
only) of the transformation.

It can be assumed that:
\begin{itemize}
\item No two classes have the same name.
\item No two types have the same name.
\item The owned attributes of each class have distinct names within the class, 
and do not have
common names with the attributes of any
superclass.
\item In this version there is no multiple inheritance, ie., the multiplicity of
$generalisation$ is restricted to
$0..1$.
\end{itemize}
These properties must also be preserved
by the transformation.

The informal transformation steps are the
following:
\begin{description}
\item[(1) Pull up common attributes of all direct subclasses:] If the set 
$g = c.specialisation.specific$ of all
direct subclasses
of a class $c$ has two or more elements,
and all classes in $g$ have an owned
attribute with the same name 
$n$ and type $t$, add an attribute of
this name and type to $c$, and remove 
the copies from each element of $g$.
% (Figure \ref{rule1fig}).
% \begin{figure}[htbp]
% \begin{center}\
% \psfig{file=cd0mm.ps,width=4.2in}
% \includegraphics[width=3.2in]{rule1fig}
% \caption{Rule 1}
% \label{rule1fig}
% \end{center}
% \end{figure}
This is the ``Pull up attribute" refactoring
of \cite{Fowl}.
\item[(2) Create subclass for duplicated attributes:] If a class $c$ has two or more 
direct subclasses 
$g$, % = c.specialisation.specific$,
and there is a subset $g1$
of $g$, of size at least 2, 
all the elements of $g1$ have an owned
attribute with the same name 
$n$ and type $t$, but there are elements of $g - g1$ without such an
attribute, introduce a new class 
$c1$ as a subclass of $c$. $c1$ should
also be set as a direct superclass
of all those classes in $g$ which own
a copy of the cloned attribute. (In order
to minimise the number of new classes
introduced, the {\em largest} set of
subclasses of $c$ which
all contain a copy of the
same attribute should be chosen).
Add an attribute of
name $n$ and type $t$
to $c1$ and remove the copies
from each of its direct subclasses.
% (Figure \ref{rule2fig}).
% \begin{figure}[htbp]
% \begin{center}\
% \psfig{file=cd0mm.ps,width=4.2in}
% \includegraphics[width=3.8in]{rule2fig}
% \caption{Rule 2}
% \label{rule2fig}
% \end{center}
% \end{figure}
This is the ``Extract superclass" refactoring
of \cite{Fowl}.
\item[(3) Create root class for duplicated attributes:] If there are two or more
root classes all of which have an owned
attribute with the same name 
$n$ and type $t$, create a new root class $c$. Make $c$ the direct superclass 
of all root classes with such an attribute,
and add an attribute of
name $n$ and type $t$
to $c$ and remove the copies from each of   the direct subclasses.
% (Figure \ref{rule3fig}).
% \begin{figure}[htbp]
% \begin{center}\
% \psfig{file=cd0mm.ps,width=4.2in}
% \includegraphics[width=3.8in]{rule3fig}
% \caption{Rule 3}
% \label{rule3fig}
% \end{center}
% \end{figure}
\end{description}
% It is a requirement of the 
% transformation 
% to minimise the number of new classes % introduced, to avoid introducing
% superfluous classes into the model.
% This means in particular
% that rule 1 
% ``Pull up attribute" should be prioritised
% over rules 2 ``Extract superclass" or 3
% ``Create root class". In addition, groups
% of cloned attributes should be removed
% in a single step, in cases 2 and 3, rather
% than (for example) creating new 
% classes for pairs of attribute clones at
% a time.

% The key tasks for this problem are:
% \begin{itemize}
% \item Specify the transformation using
% a particular model transformation
% language.
% \item Show that the transformation 
% establishes
% the required properties of the target
% model (the properties to be preserved,
% listed above).
% \item Define an implementation of the
% transformation, and evaluate the
% effectiveness and quality
% of this implementation on
% the test cases, using the criteria of
% Section \ref{evalcrit}.
% \item Analyse the confluence and
% termination properties of this 
% implementation. 
% \item Evaluate the efficiency and 
% maximum capability of the 
% implementation, using a common
% execution platform, the SHARE
% platform.
% \item Identify the syntactic complexity
% and modularity of the solution.
% \item Identify (by means of a survey)
% the understandability of the solution.
% \end{itemize} 

\subsection{Test cases}
\label{testcases}

The solutions should be tested on
the following three
test cases of increasing size
and complexity. These test cases 
represent both typical scenarios
which could be expected to arise
in class diagram modelling
(test cases 1 and 2), and pathological
examples designed to check the
behaviour of the transformation in
extreme cases (test case 3 and the 
duplications of test case 2). 
% The test cases are provided in
% Eclipse XSI XML format.
% The test models can be found on the
% case SHARE VM, under `ttc13refactoring'.

% Files testcase1.xsi.txt, testcase2.xsi.txt, % etc, contain the input models in XSI 
% format, together with plain text data
% equivalents testcase1.txt, etc. The 
% results are in files test1result.xsi.txt, etc.
%  A file mm.txt (for UML-RSDS) and 
% My.ecore (for Eclipse) defines the
% metamodel. Alternative formats of the
% files and metamodel are in the 
% $model$ and $metamodel$
% subdirectories.

\begin{figure}[htbp]
\begin{center}\
\includegraphics[width=3.0in]{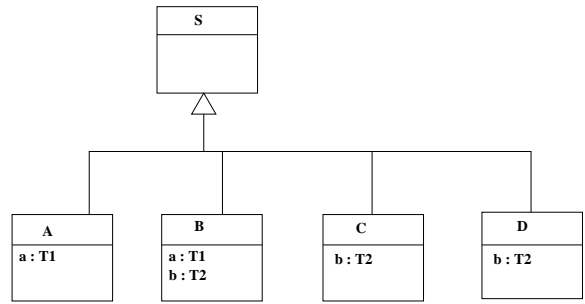}
\caption{Test case 1}
\label{testcase1}
\end{center}
\end{figure}

The first test case is a simple test for
alternative applications of rule 2.
Figure \ref{testcase1} shows the 
starting model.

Applying the rule to classes B, C, D to
remove duplicates of b is the
preferred choice because it creates 
fewer new classes
than an application to A and B, to
remove the duplicate of a, followed
by an application to C and D,
although
both solutions remove the maximum
possible 2 clone attribute copies.

% The resulting model should therefore
% have a superclass of $B$, $C$ and $D$
% containing an attribute with name 
% ``b" and type $T2$ (Figure
% \ref{testcase1result}).
% \begin{figure}[htbp]
% \begin{center}\
% % \psfig{file=testcase2.ps,width=4.2in}
% \includegraphics[width=3.0in]{testcase1result}
% \caption{Test case 1 result}
% \label{testcase1result}
% \end{center}
% \end{figure}

\begin{figure}[htbp]
\begin{center}\
\includegraphics[width=3.5in]{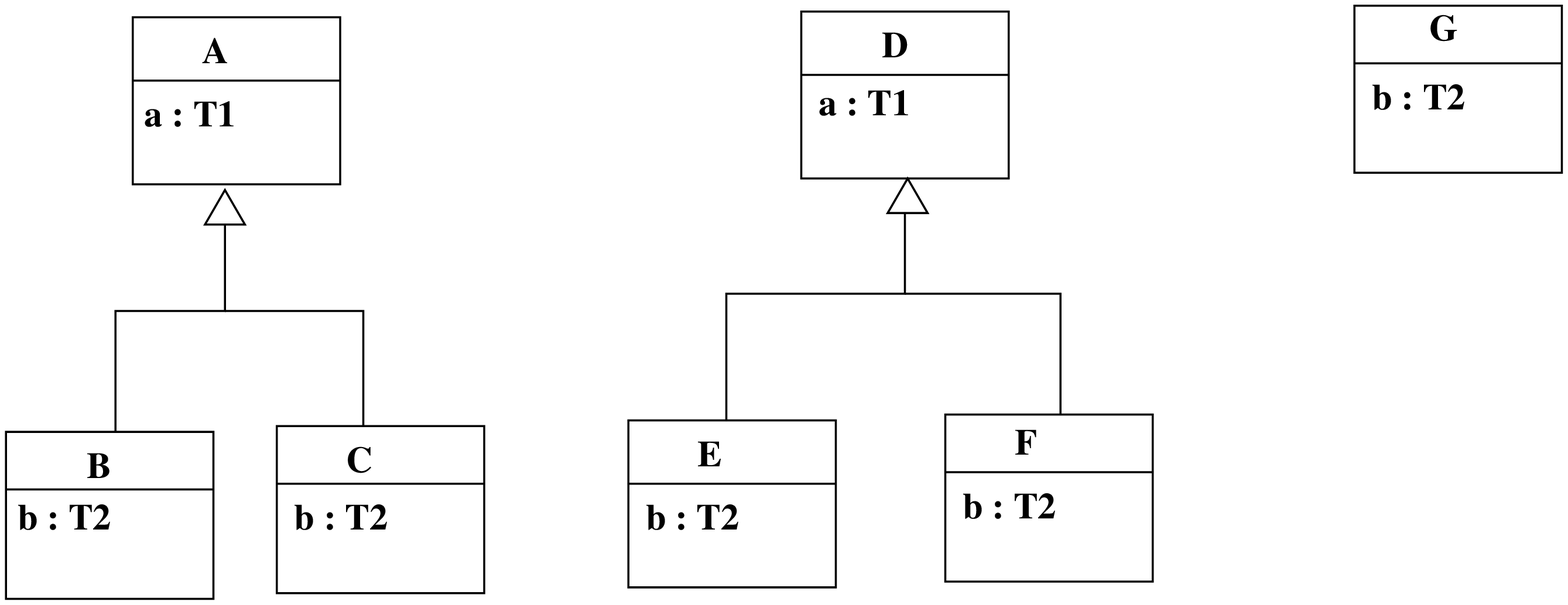}
\caption{Test case 2}
\label{testcase2}
\end{center}
\end{figure}

A larger test case, involving applications of rules 1 ``Pull up attributes"
and 3 ``Create root class", 
is shown in Figure 
\ref{testcase2}. 

% The ideal result of applying the
% transformation to this test is shown in
% Figure \ref{testcase2res}, but other 
% results are possible in which the 
% factoring
% of duplicated attributes is incomplete. 
% For example, the sequence of rule
% applications:
% \begin{verbatim}
% b from E, F to D   (rule 1)
% b from D, G to DG  (rule 3)
% b from B, C to A  (rule 1)
% b from A, DG to ADG  (rule 3)
% \end{verbatim}
% fails to amalgamate the clone copies of
% attribute $a$, and has 80\% 
% effectiveness (of 5 possible clone copies % that can be
% removed, only 4 are removed).

% \begin{figure}[htbp]
% \begin{center}\
% \includegraphics[width=3.5in]{testcase2res}
% \psfig{file=testcase2res.ps,width=4.2in}
% \caption{Expected result of test 2}
% \label{testcase2res}
% \end{center}
% \end{figure}

Test case 3 has 500 classes, each of
which is a root class, and there are
ten attributes in each class, with the
attributes of each class being a copy of
those in each other class (ie., 5000 
attributes, with 4990 clone copies).
Only one new class needs to be introduced,
as a superclass of all the other classes,
and all redundant copies of the attributes
can be removed.

In addition we recommend carrying
out `stress
testing' to measure the maximum 
capability of a transformation tool and
implementation, 
in terms of the maximum size of
models which a transformation
tool or implementation is capable of 
processing. These tests are
models formed
from duplicated copies of test case 2
(omitting class $D$ and its subclasses),
of sizes up to 10000 copies (40000 
classes, 40000 attributes, 20000
generalisations).

Table \ref{testcasesumm} summarises
the test cases.

\begin{table}[htbp]
\centering
\begin{small}
\begin{tabular}{l|l|l|l}
{\em Test case} & {\em Number of} & {\em Number of} & {\em Total} \\
                 & {\em classes} & {\em attributes} & {\em size}  \\ \hline
1 & 5 & 5 & 14 \\
2 & 7 & 7 & 18  \\
% 3 & 105 & 132 & 337  \\
3 & 500 & 5000 & 5500  \\
2*1000 & 4000 & 4000 & 10000   \\
2*5000 & 20000 & 20000 & 50000  \\
2*10000 & 40000 & 40000 & 100000   \\
\end{tabular}
\caption{Test cases}
\label{testcasesumm}
\end{small}
\end{table}

Solution providers should ensure that
their solutions can successfully process
test cases 1, 2 and 3 at a minimum, and
as many of the capability test cases as
possible. Models should be processed
as XMI files, eg., as exported by 
Eclipse. Self-evaluation of the 
solutions using the evaluation 
criteria in Appendix \ref{evalcrit}
should be provided, in addition 
a survey of usability should be carried
out among other TTC participants, at
least 5 persons not involved in 
creating the solution.

\section{Extension: adaption to work with multiple inheritance}
\label{assocclones}

In this extension the transformation 
should be generalised to work also
on models containing multiple
inheritance. This implies alternative
strategies for the three rules of  
Section \ref{attclones}:
\begin{enumerate}
\item For rule 1, if there are multiple
superclasses $s1$, ..., $sn$ each with
a common set $c1$, ..., $cm$ of subclasses
with an owned attribute of the same
name and type, should the common
attribute be moved up to all of the
$si$ (thereby 
introducing name clashes in the
$cj$), or to only one of them, and in that
case, which one?
\item For rules 2 and 3, there is now the
possibility to group together all classes
which contain a common attribute, as
subclasses of a new class, even if there
are overlapping groups. Introducing
multiple inheritance may complicate
the structure of the class diagram, and
probably should not be used in every 
such case.
\end{enumerate}

The effort required to modify the case
study solution to meet the extended
requirement should be recorded as an
evaluation measure of
extensibility.

\providecommand{\urlalt}[2]{\href{#1}{#2}} 
\providecommand{\doi}[1]{doi:\urlalt{http://dx.doi.org/#1}{#1}}

\appendix 

\section{Evaluation criteria}
\label{evalcrit}

As the basis of a systematic evaluation
framework for model transformations,
we propose to use the International Organisation for Standardization (ISO) 
standards related to  software quality, specifically the 
ISO/IEC 9126-1 standard, 
which is based upon the 
definition of a \emph{Quality Model} and its use for software evaluation
\cite{iso9126}. This framework 
defines quality models based on 
general characteristics of software, which are further refined into subcharacteristics. % Table \ref{tab:ISOIEC91261} 
% enumerates the six quality 
% characteristics defined in 
% ISO/IEC 9126-1 and their decomposition % into subcharacteristics. 

% \begin{table}[htbp]
% 	\centering
% \begin{small}
% \begin{tabular}{|l|l|}
% \hline
% Characteristics & Subcharacteristics  \\ % \hline
% {Functionality}  & Suitability, Accuracy, \\
%    & Interoperability, Security \\
%    & Functionality compliance \\ \hline
% {Reliability}  & Maturity, Fault tolerance \\
%    & Recoverability \\
%    & Reliability compliance  \\ \hline
% {Usability}  & Understandability, Learnability \\
%  & Operability, Attractiveness \\
%   & Usability compliance \\ \hline
% {Efficiency}  & Time behavior \\
%   & Resource utilisation \\
%   &  Efficiency compliance\\ \hline  
% {Maintainability}  & Analysability, Changeability \\
%   & Stability, Testability \\
%  & Maintainability compliance \\ \hline  
% {Portability}  & Adaptability, Installability \\
%   & Co-existence, Replaceability \\
%   & Portability compliance \\ \hline
%   \end{tabular}
% \caption{ISO/IEC 9126-1 quality characteristics}
% \label{tab:ISOIEC91261}
% \end{small}
% \end{table}

Relevant characteristics and subcharacteristics for evaluation of model transformation can be selected from the
ISO/IEC9126-1 framework. These
characteristics and subcharacteristics 
can then be further decomposed
into measurable attributes. 
Table \ref{tab:ISOIEC91261MT} summarizes the chosen characteristics, subcharacteristics and their
corresponding measurable
attributes.  One attribute may be related to more than one quality factor.

\begin{table}[htbp]
	\centering
\begin{tabular}{|l|l|l|} \hline
{Characteristic} & Subcharacteristic  & Attribute \\  \hline
{Functionality} & Suitability  & Abstraction level \\
&  & Size \\ 
&  & Complexity \\ 
&  & Effectiveness \\ 
&  & Development effort\\
&  & Execution time\\ \cline{2-3}
& Accuracy & Correctness \\
 &  & Completeness \\ \cline{2-3}
& Interoperability & Embeddable in transformation process \\
  & & Close to well-known notation\\
& & Interoperable with Eclipse \\ \cline{2-3}
& Functionality & Close to well-known\\
& compliance & notation \\ \hline
Reliability & Maturity & History of use\\ \cline{2-3}
  & Fault tolerance & Tolerance of false assumptions\\ \hline
{Usability} & Understandability & Survey \\
  & Learnability & results\\
   & Attractiveness & \\  \hline
{Efficiency} & Time behavior & Execution time \\
&  & Maximum capability \\
 \hline
{Maintainability} & Changeability & Size \\
&  & Complexity\\
&  & Modularity\\ \hline
{Portability} & Adaptability & Extensibility \\
 \hline 
\end{tabular}
\caption{Selected quality characteristics for evaluation of model transformation approaches}
\label{tab:ISOIEC91261MT}
\end{table}

% `

Characteristics such as interoperability
and adaptability can be interpreted and
evaluated in several different ways.
Here we have evaluated these 
characteristics based upon key factors
specific to model transformations. For
example, the ability to interwork with
Eclipse is an important factor for the interoperability
of a transformation approach.

In cases where a numeric value is not
appropriate for an attribute
(such as Abstraction level, Maturity)
a three-point or five-point scale is
used to summarise the relative
values of attributes.

The following are the specific measures
which should be evaluated for each
solution:
\begin{itemize}
\item Size: lines of specification text
\item Complexity: sum of number of 
operator occurrences and feature and
entity type name references in the 
specification expressions
\item Effectiveness: proportion of 
attribute clones which are removed,
relative to the theoretical maximum
number that can be removed
\item Development effort: developer time
in person-hours spent in writing and
debugging the specification
\item Execution time: milliseconds 
\item History of use: number of years 
the model transformation language and tools have been publicly
available
\item Maximum capability: maximum
size of input model, in terms of 
number of elements, which can be 
successfully processed
\item Modularity: proportion of calls
internal to modules, relative to the
total number of calls. Value is 1 if 
there are no external calls.
\end{itemize}
 
Abstraction level is classified as High 
for primarily declarative solutions,
Medium for declarative-imperative
solutions, and Low for primarily
imperative solutions.

The effectiveness measure used
is the
proportion of clone copies of attributes
which are removed by the transformation. That is, if there are $n$ copies of attributes
which could, in principle, be removed by
the rules 1, 2 and 3, and the
implemented transformation removes
$m \leq n$ copies, the effectiveness is
$m/n$.
In addition, a solution is optimal if the
minimum possible
number of new classes are introduced.

% \begin{table}[htbp]
% \centering
% \begin{tabular}{|l|l|l|l|}
% + & {\em Low} & {\em Medium} & {\em High} \\ \hline
% {\em Low} & Low & Medium & High \\
% {\em Medium} & Medium & High & High \\
% {\em High} & High & High & High \\ 
% \end{tabular}
% \caption{Sum of complexity measures}
% \label{sumcomp}
% \end{table}

Execution time of the transformation
implementation does not include the
loading and unloading of models from
the transformation tool. 

% Maximum capability is measured as
% the maximum 
% number of elements in a model
% which can be processed
% within 30 minutes.

Correctness is divided into syntactic
correctness, termination and confluence.
Syntactic correctness is 
the capability to establish 
the constraints of the target
metamodel of a transformation, and
the capability to 
establish or preserve
correct inverse links to associations
(in this case study the pairs $general$/$specialisation$ and
$specific$/$generalisation$ of roles).
Equivalently, it is the ability to ensure
conformance of the target model to
the target metamodel.
The classification of correctness is given
by an average of three separate 5-point
measures for syntactic correctness, 
termination and confluence.
% (Table \ref{correctnesstable}). 
Each measure separately is
rated -2 (None), -1 (Low), 0 (Medium), 1 (High) and 2 (Comprehensive).

Usability is separated into:
\begin{itemize}
	\item  Understandability: how easy a transformation specification is to 
comprehend. 
\item Learnability: the degree to
which the transformation language and
tool can be learnt in a reasonable
timescale and with reasonable effort. 
% Learnability is related to 
% Understandability. The easier a 
% language and tool is to understand, the % easier it is to learn and relearn them. 
\item Attractiveness: how acceptable
is the language and tool for the user. 
\end{itemize}

Empirical studies with representative
users and tasks are considered one of the best techniques to measure usability of
software systems. Therefore we 
propose that a survey
of TTC participants (at least 5 persons
not involved in the surveyed case study
solution)
is used to measure the Understandability,  Learnability and Attractiveness of 
solutions to the case study. 
The survey 
contains five questions, each with
5 answer options from None (0), Low (1),
Medium (2), High (3), Very high (4). 
The first question identifies the level of 
knowledge of the specific transformation 
language by the participant. 
The rest of the
questions are as follows:
\begin{itemize}
\item To what degree do
the rules of the transformation satisfy 
the case study description / How easy is it
to relate the informal to the
formal specification? (Understandability)
\item How well structured is the transformation specification? (Attractiveness)
\item How attractive is the 
specification notation to
read? (Attractiveness)
\item How much effort is needed to understand the transformation? (Learnability)
\end{itemize}
In addition to 
these questions we include a small test case to assess the actual
understanding of the transformation
by a participant: the participant needs 
to explain where in the transformation
specification a 
particular aspect of the transformation
(promoting a 
duplicated attribute to
a superclass) is dealt with. This is a 
factor for understandability.

The difference between this score for
detailed understanding and the level of 
initial knowledge is also taken as
a learnability factor.

Interoperability
consists of Embeddability: how 
effectively the transformation
can be reused within a 
larger quality improvement process,
consisting of transformations to
(1) remove redundant inheritance,
(2) remove multiple inheritance,
(3) replace concrete superclasses by an
abstract class and a new concrete
subclass of this class. Embeddability is
High if both internal and external
composition of these transformations is
possible, Medium if only one composition
is possible, and Low if no composition is
possible.

Another factor for interoperability is
the closeness to a well-known notation,
which is graded 
by a three-point scale: High (+1)
for a common syntax and semantics to
a well-known notation (eg., OCL);
Medium (0) for a variant syntax and/or 
variant semantics; Low (-1) for no 
similarity.

Interoperability with Eclipse is given by
a three-point scale: High (+1) for complete
integration; Medium (0) for interoperability
via exported/imported data files only;
Low (-1) for no interoperation 
mechanism.

Maturity is considered low (-1) for 
languages/tools of less than 4 years
public availability, medium (0) for 4 
up to eight years availability, 
and high (+1) for more than 8 years.

Extensibility is measured by evaluating
how succesfully the solution for the core
problem can be generalised to solve the
extended case study problem in
Section \ref{assocclones}. It is measured
quantitatively by the effort required to
extend the solution, in person-hours.
 
\section{UML-RSDS Solution}

The case study transformation
has been specified in UML-RSDS
\cite{umlrsds}. This
specification
consists of the class diagram of
Figure \ref{cd0mm}, 
and a single use case 
which represents the transformation.
The use case has precondition
constraints
expressing the assumptions  
of the transformation, and a sequence
 of
three postcondition constraints 
$(C1)$, $(C2)$, $(C3)$ 
corresponding to the three informal
rules. Each of these operates on instances
of $Entity$:

\[ (C1):\\
a : specialisation.specific.ownedAttribute ~\&~ \\
specialisation.size > 1 ~\&~ \\
specialisation.specific{\fun}forAll(\\ 
\t1 ownedAttribute{\fun}exists( b | b.name = a.name ~\&~ b.type = a.type ) ) ~~\implies\\
\t2 a : ownedAttribute  ~\&\\
\t2 specialisation.specific.ownedAttribute{\fun}select(\\
\t8 name = a.name ){\fun}isDeleted()\]

This specifies that an  
instance  ($self$) of $Entity$, and
instance $a$ of $Property$ match the
constraint LHS if:
(i) $a$ is in the set of attributes of all
direct subclasses of $self$, (ii) there is
more than one direct subclass of
$self$, and (iii) every direct
subclass of $self$ 
has an attribute with the same name and
type as $a$.

The conclusion specifies that (i) the
property $a$ is moved up to the
superclass $self$, (ii) all other attributes
with name $a.name$ are deleted from
all direct subclasses of $self$. 

$s{\fun}isDeleted()$ is a built-in operator
of UML-RSDS, which deletes the object or
set of objects $s$ from their model, 
removing them from all entity types and
association ends.
 
\[ (C2):\\
a : specialisation.specific.ownedAttribute ~\&~ \\
v = specialisation{\fun}select(\\ 
\t1  specific.ownedAttribute{\fun}exists( b | b.name = a.name ~\&~ b.type = a.type ) ) ~\&\\
v.size > 1 ~\& \\
specialisation.specific.ownedAttribute{\fun}forAll( c | specialisation{\fun}select(\\ 
\t1  specific.ownedAttribute{\fun}exists( d | \\
\t2 d.name = c.name ~\&~ d.type = c.type ) ){\fun}size() ~\leq~ v.size ) ~~\implies\\
\t1 Entity{\fun}exists( e | e.name = name + ``\_2\_" + a.name ~ \&\\
\t2  a : e.ownedAttribute  ~\&\\
\t2 e.specialisation = v ~\&\\
\t2 Generalization{\fun}exists( g | g : specialisation ~\&~ g.specific = e ) ) ~\&\\
\t2 v.specific.ownedAttribute{\fun}select( name = a.name ){\fun}isDeleted() 
\]

The assumption specifies that an  
instance  ($self$) of $Entity$, and
instance $a$ of $Property$ match the
constraint LHS if:
(i) $a$ is in the set of attributes of all
direct subclasses of $self$, (ii) the set
$v$ of all specialisations of $self$ whose
class contains a clone attribute of $a$
has size greater than 1, (iii) $v$ is of
maximal size over all groups of
specialisations of $self$ which contain a
common attribute.
 
The conclusion specifies that: (i) a new
entity $e$ is created, and the
property $a$ is moved up to $e$, (ii)
the specialisations of $e$ are $v$, 
(iii) $e$ is made a subclass of $self$, and
(iv) all the clone copies of $a$ in 
$v$ are deleted.

\[ (C3): \\
a : ownedAttribute  ~\&\\
generalisation.size = 0 ~\&~\\ 
v = Entity{\fun}select( generalisation.size = 0 ~\&~\\
\t1 ownedAttribute{\fun}exists( b | b.name = a.name ~\&~ b.type = a.type ) ) ~\&\\
v.size > 1 ~~\implies\\
\t1 Entity{\fun}exists( e | e.name = name + ``\_3\_" + a.name ~\&~\\ 
\t2  a : e.ownedAttribute  ~\&\\
\t2  v.ownedAttribute{\fun}select( name = a.name ){\fun}isDeleted() ~\&\\
\t2  v{\fun}forAll( c | Generalization{\fun}exists(\\
\t4 g | g : e.specialisation ~\&~ g.specific = c ) ) ) \]

The Java executable of this solution is
provided in the Share VM, as GUI.java.
This can be executed by the command
{\tt java GUI} from the command line. The
executable reads its input model from
in.txt
and produces output in the file out.txt,
when the button for the transformation
is selected.
XMI output is also produced, in xsi.txt.


\begin{thebibliography}{99}

\bibitem{iso9126}
    {Botella, P., Burgu\'{e}s, X., Carvallo, J. P., Franch, X., Grau, G., Marco, J., Quer, C.},
   {\em {ISO/IEC 9126 in practice: what do we need to know?}},
{Software Measurement European Forum (SMEF 2004)}.   

\bibitem{Fowl} M. Fowler, K. Beck,
J. Brant, W. Opdyke, D. Roberts,
{\em Refactoring: improving the design of existing code}, Addison-Wesley, 1999.

\bibitem{Lano12scp}
S. Kolahdouz-Rahimi, K. Lano, 
S. Pillay, J. Troya, P. Van Gorp,
{\em Goal-oriented measurement of model 
transformation methods}, Science of
Computer Programming, 2013, \doi{10.1016/j.scico.2013.07.013}.

\bibitem{umlrsds} K. Lano,
{\em The UML-RSDS Manual},
\url{http://www.dcs.kcl.ac.uk/staff/kcl/uml2web}, 2013.

\bibitem{Pere10} J. Perez, Y. Crespo,
B. Hoffmann, T. Mens,
{\em A case study to evaluate the 
suitability of graph transformation
tools for program refactoring},
Int. J. Softw. Tools Technical Transfer (2010), vol 12: 183--199,
\doi{10.1007/s10009-010-0153-y}.

\end{thebibliography}
\end{document}